\documentclass{emulateapj}

\newcommand{\Ms}{\,\mathrm{M_\odot}} 					
\newcommand{\keV}{\textrm{keV}}
\newcommand{\Mpc}{\textrm{Mpc}}
\newcommand{\kpc}{\textrm{kpc}}
\newcommand{\cm}{\textrm{cm}}

\newcommand{\ks}{\mathrm{ks}}
\newcommand\Chandra{%
  {\it Chandra}}
\newcommand\XMM{%
  {\it XMM-Newton}}

\newcommand\eqnref[1]{%
 Eqn.~\ref{eqn:#1}}

\newcommand\tabref[1]{%
Tab.~\ref{tab:#1}}

\newcommand\figref[1]{%
Fig.~\ref{fig:#1}}

\newcommand\secref[1]{%
Sec.~\ref{sec:#1}}

\begin{document}

\shortauthors{Riemer--S{\o}rensen et al.}
\title{Resolving the discrepancy between lensing and X-ray mass estimates of the complex galaxy cluster Abell~1689}
\shorttitle{Lensing and X-ray mass estimates of Abell~1689}

\author{S. Riemer--S{\o}rensen\altaffilmark{1}, D. Paraficz\altaffilmark{1}, D.D.M. Ferreira\altaffilmark{1}, K. Pedersen\altaffilmark{1},
M. Limousin\altaffilmark{1,2}, H. Dahle\altaffilmark{3,4}}
\altaffiltext{1}{Dark Cosmology Centre, Niels Bohr Institute, University of Copenhagen,
Juliane Maries Vej 30, DK-2100 Copenhagen, Denmark}
\altaffiltext{2}{Laboratoire d'Astrophysique de Toulouse-Tarbes, Universit\'{e} de Toulouse, CNRS, 57 avenue d'Azereix, F-65000 Tarbes, France}
\altaffiltext{3}{Institute of Theoretical Astrophysics, University of Oslo, P.O. Box 1029, Blindern, N-0315 Oslo, Norway}
\altaffiltext{4}{Centre of Mathematics for Applications, University of Oslo, P.O. Box 1053 Blindern, N-0316 Oslo, Norway}
\email{signe, danutas, kp, desiree, kp@dark-cosmology.dk, marceau.limousin@ast.obs-mip.fr, hakon.dahle@astro.uio.no}

\begin{abstract}

There is a long-standing discrepancy between galaxy cluster masses determined from X-ray and gravitational lensing observations of which Abell~1689 is a well-studied example. In this work we take advantage of 180~ks of \Chandra{} X-ray observations and a new weak gravitational study based on a Hubble Space Telescope mosaic covering the central $1.8~\Mpc \times 1.4~\Mpc$ to eliminate the mass discrepancy. In contrast to earlier X-ray analyses where the very circular surface brightness has been inferred as Abell~1689 being spherically symmetric and in hydrostatic equilibrium, a hardness ratio map analysis reveals a regular and symmetric appearing main clump with a cool core plus some substructure in the North Eastern part of the cluster. The gravitational lensing mass model supports the interpretation of Abell~1689 being composed of a main clump, which is possibly a virialized cluster, plus some substructure. In order to avoid complications and mis-interpretations due to X-ray emission from the substructure, we exclude it from the mass reconstruction. Comparing X-ray and lensing mass profiles of the regular main part only, shows no significant discrepancy between the two methods and the obtained mass profiles are consistent over the full range where the mass can be reconstructed from X-rays (out to $\approx 1 \Mpc$). The obtained cluster mass within $\approx875$~\kpc{} derived from X-rays alone is $6.4\pm2.1\times10^{14}\Ms$ compared to a weak lensing mass of $8.6\pm3.0\times10^{14}\Ms$ within the same radius.
\end{abstract}


\section{Introduction}
Gravitational lensing and X-ray observations are two independent methods, which can probe the mass of the largest virialized structures in the Universe, namely clusters of galaxies. The cluster mass is dominated by dark matter ($\approx 80\%$) and hot X-ray emitting gas ($\approx 20\%$). The baryonic matter in the galaxies only contributes a few percent of the mass. The spectrum of the X-ray emission from the gas depends on the temperature of the gas, which for a cluster in hydrostatic equilibrium is related to the total mass of the cluster. Gravitational lensing uses the distortion of background source images to probe the total mass along the line of sight. It has been widely debated whether cluster mass determinations from the two methods agree, and in some cases large discrepancies have been found \citep[e.g.][]{Loeb:1994,Miralda:1995,Voigt:2006,Mahdavi:2007sd,Zhang:2008tr}. The main motivation for studying a single cluster, Abell~1689, in large detail with the best available X-ray and lensing data is to get insight into the mass discrepancy by determining the mass distribution using both methods. It is checked that the results of a third independent mass determination method using the velocity dispersion of the galaxies within the cluster are consistent with the X-ray and lensing results.

Abell~1689 is a massive cluster with a redshift of $z=0.1832$ \citep{Teague:1990}. It is mostly known for its amazing gravitational arcs and large number of multiply imaged systems. Consequently it has been well studied with strong and weak gravitational lensing \citep{Broadhurst:2004,Broadhurst:2005,Halkola:2007,Lemze:2007,Limousin:2007,Dahle:2008}. It has been proposed as a standard example of a spherical cluster of galaxies in hydrostatic equilibrium \citep{Xue:2002,Lemze:2007}. However, earlier studies have found large discrepancies between the mass obtained from X-ray observations and from gravitational lensing \citep{Andersson:2004,Andersson:2006,Lemze:2007}. Strong gravitational lensing analyses have found the central 300~\kpc{} to consist of several subclumps \citep{Limousin:2007,Tu:2007} and weak lensing reveals substructure on a larger scale (500~\kpc{}) \citep{Okura:2008,Dahle:2008}. Furthermore an indication of substructure in the North Eastern (NE) part of Abell~1689 was seen by \citet{Andersson:2004} in an \XMM{} X-ray observation.

Recently new data have become available in the form of a very deep X-ray observation with \Chandra{} ($\approx 150$~ks new exposure). In addition, a weak gravitational lensing analysis based on a mosaic using the Wide Field Planetary Camera 2 (WFPC2) instrument aboard the Hubble Space Telescope (HST) has recently been carried out \citep{Dahle:2008}. This provides a unique mass map with an unprecedented combination of spatial resolution ($50\arcsec$) and a large spatial extension. We explore both of these data sets in this work.

The long X-ray exposure allows us to create hardness ratio maps with a resolution of 5\arcsec{} (\secref{analysis_HR}), revealing substructure in the X-ray emission from the NE part of the cluster. In contrast, the South Western (SW) part seems circular and very regular. Consequently we assume it to be spherically symmetric and relaxed, and we reconstruct the temperature and mass profiles from SW part alone (\secref{analysis_temp}). The mass profile of the SW part is compared to the lensing mass profile of the same region (\secref{result_lensing}). The obtained profiles are in very good agreement and it is proposed that Abell~1689 consists of a spherically symmetric part plus some substructure in the NE part (\secref{discussion_xray}). We conclude that the mass profiles determined from high quality X-ray and gravitational lensing data are in agreement.

Throughout the paper we have used a standard $\Lambda$CDM cosmology with $\Omega_{baryon}=0.046$, $\Omega_{\Lambda}=0.72$, $\Omega_{CDM}=0.23$, and $H_0=70.1\,$km$\,$s$^{-1}\,$Mpc$^{-1}$ \citep{wmap5yr}. At the redshift of Abell~1689, 1\arcmin{} corresponds to 185~\kpc{} for the chosen cosmological model. Unless otherwise stated, quoted uncertainties are one $\sigma$.

\section{X-rays}
\subsection{X-ray observations}
We have analyzed four \Chandra{} X-ray observations from the NASA HEASARC archive\footnote{\url{http://heasarc.gsfc.nasa.gov/docs/archive.html}} with a total exposure of approx. 180~ks (see \tabref{obs}). It has been claimed that the two newest, and by far longest, observations cannot be used due to background issues \citep{Lemze:2007}, but these can be overcome, as shown in \secref{background}. 

\begin{deluxetable}{c|c|c|l}
\tablecaption{\label{tab:obs}The analysed \Chandra{} observations of Abell~1689.}
\startdata
Observation id 	& Date 		& Exposure time\tablenotemark{a} & Data mode \\ \hline
1663			& 2001 Jan 7	& 10.73~ks	& FAINT\\
5004			& 2004 Feb 28	& 19.86~ks		& VFAINT\\ 
6930			& 2006 Mar 06	& 75.79~ks		& VFAINT\\
7289			& 2006 Mar 09	& 74.61~ks		&  VFAINT
\enddata
\tablenotetext{a}{The given exposure times are after light curve cleaning.}
\end{deluxetable}

The four observations in \tabref{obs} have been reprocessed with the calibration CALDB 3.4.1 and analysed with CIAO 3.4 \citep{Fruscione:2006} following standard procedures\footnote{\url{http://cxc.harvard.edu/ciao3.4/}, The CIAO Data Analysis Page}. We found the X-ray peak of Abell~1689 to have the position (R.A., decl.)=(197.87306$^o$, -1.3413889$^o$). In all observations, the X-ray peak was situated on the ACIS-I3 chip and we performed the analysis with events entirely from this chip.

\subsection{X-ray analysis: Image and surface brightness}
\begin{figure}[tbp]
\centering
	\includegraphics[width=8.5cm]{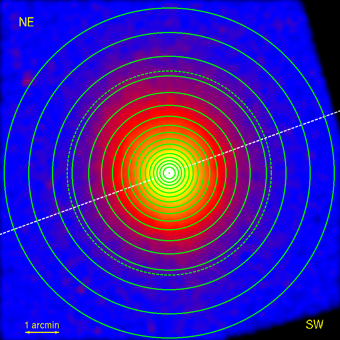}
	\caption{Exposure corrected X-ray image composed from the observations in \tabref{obs} (smoothed with a $10\arcsec$ wide Gaussian). The dashed white line through the X-ray peak (20$^o$ angle) divides the cluster in the NE and SW halves. The annuli described in \secref{analysis_temp} are shown in solid green and the $3\arcmin$ circle in dashed green.}
	\label{fig:xray}
\end{figure}

\begin{figure}[tbp]
\centering
	\includegraphics[width=8.5cm]{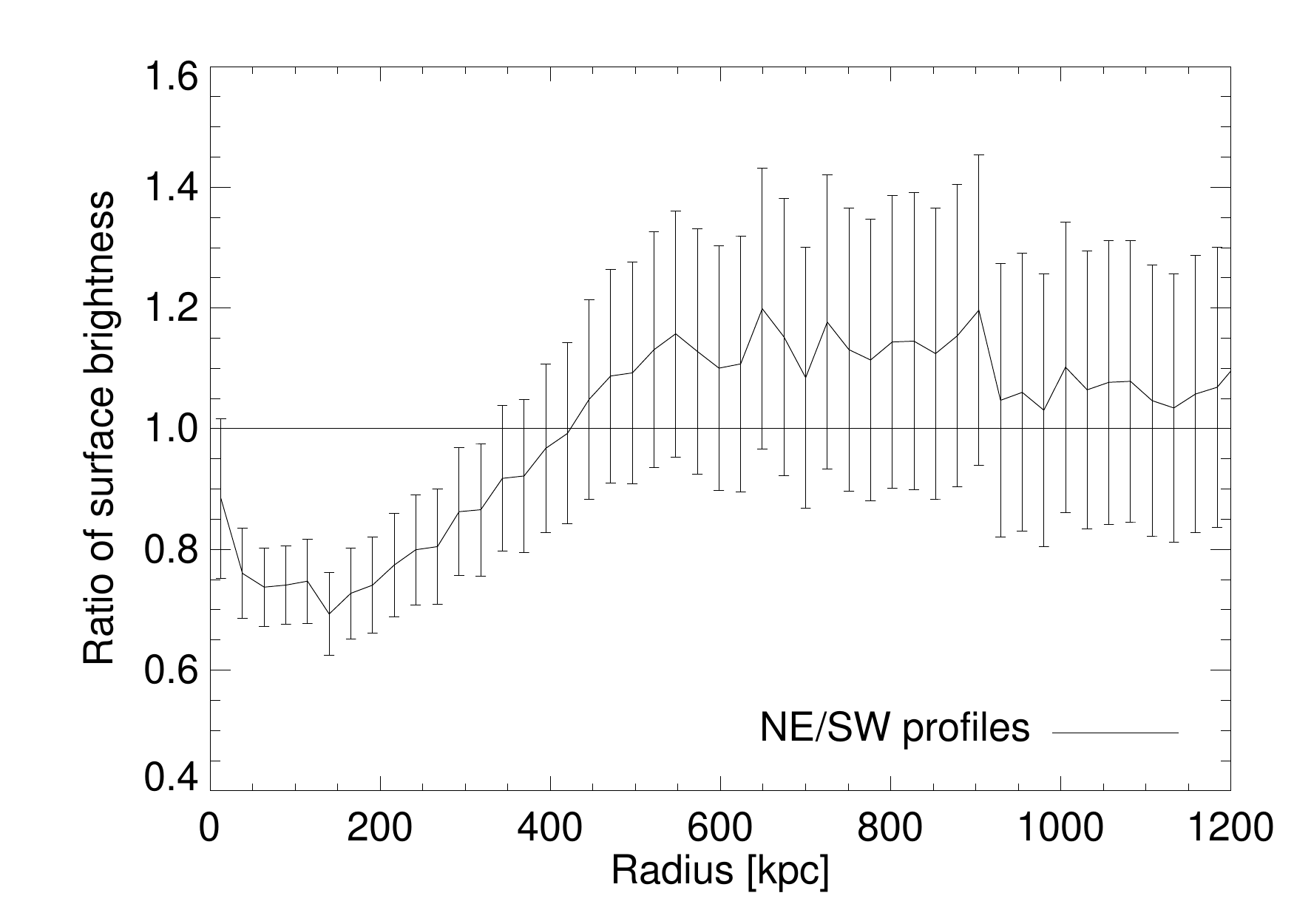}
	\caption{The ratio between the surface brightness profiles in the $0.3-12.0$~\keV{} of the NE and SW halves with Poisson errors.}
	\label{fig:ratio}
\end{figure}

An exposure corrected X-ray image combined from the four observations is shown in \figref{xray}. The emission is almost circular tempting one to conclude that Abell~1689 is spherical and in hydrostatic equilibrium. However this is not the case, which is clearly demonstrated by the surface brightness profiles derived from \figref{xray} for the NE and SW halves respectively (defined by the dashed white line). \figref{ratio} shows the ratio between the NE and SW profiles. It is evident that within $\approx 300$~\kpc{} the SW half is brighter, where outside $\approx 500$~\kpc{} the two halves are more equally bright with a tendency for the NW half to be brighter.

\subsection{X-ray analysis: Background} \label{sec:background}
\Chandra{} has two telemetry modes called FAINT and VFAINT (Very FAINT). They have different event gradings and thereby different background event selections \citep[ch. 6.14]{POG9}. Three of the four observations analysed here have been done in the VFAINT mode.
The standard blank sky observations for background subtraction are produced from observations in the FAINT mode \citep{Markevitch:blank}, and hence the shape of the background spectrum from a VFAINT observation deviates from the spectrum of a blank sky observation. However, instead of using the pipeline processed VFAINT data, the data can be reprocessed as FAINT observations. In the latter case, the shape of the background spectra of the two new, long observations 6930 and 7289 are identical to the corresponding blank sky spectra. We have extracted the blank sky spectra from the same chip regions as the observed cluster spectra and scaled the blank sky spectrum level to the corresponding observational spectrum in the 9-12~\keV{} interval, where very little cluster emission is expected. 

Instead of using the blank sky background, it is possible to use the local background from the same observation allowing to take advantage of the improved background reduction in the VFAINT mode. The background differs from chip to chip (in the CCD), so the local background has to be extracted from the same chip as the source region (in this case ACIS-I3). We have compared the two methods of background subtraction and the difference between the final results (e.g. the temperature) is negligible. In the following we have used the blank sky method whenever extracting spectra, since it allows us to use more of the chip for the actual analysis and hence determine the temperature further from the centre of the cluster. For the images we have used the VFAINT data since the background event grading is better.

\subsection{X-ray analysis and results: Hardness ratio} \label{sec:analysis_HR} \label{sec:result_HR}
The ratio of low energy photons to high energy photons from the intra cluster medium is a proxy of the temperature structure of the cluster. If the cluster is isothermal and only radiates thermally, the ratio between soft and hard photons (S/H) will be independent of position (within the cluster). However, if there is a variation in temperature, we expect to see a difference in S/H. Even clusters in hydrostatic equilibrium have a radial temperature gradient, which we expect to see as a circular structure in the soft to hard photon ratio.

\begin{figure}[tbp]
\centering
	\includegraphics[width=8.5cm]{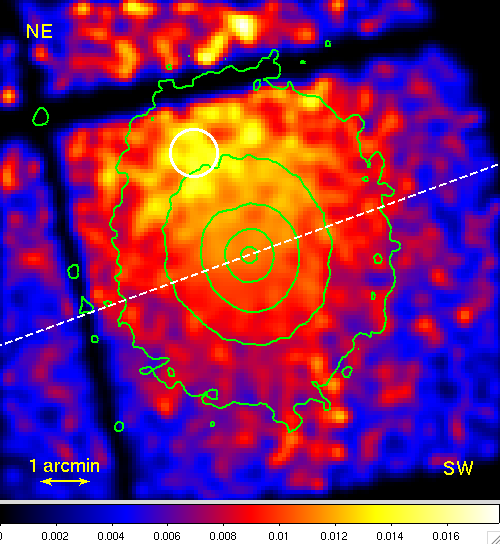}
	\includegraphics[width=8.5cm]{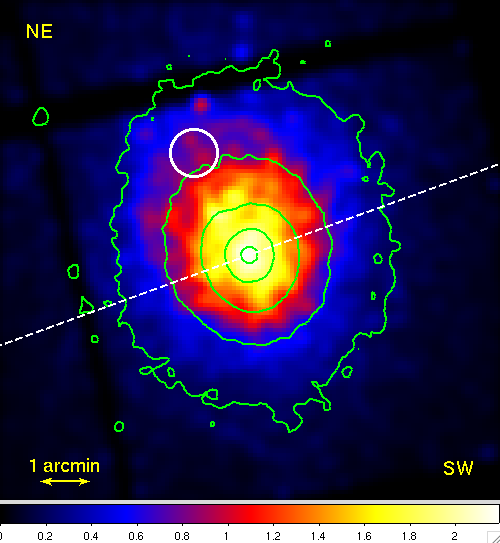}
	\caption{The obtained hardness ratio maps (bright is soft photon excess). The green contours show the almost circular total X-ray emission. The white line divides the cluster into the NE (upper) part and the SW (lower) regular and symmetrically appearing half. The white circle is the region analysed in \secref{analysis_NE}. {\it Upper:} The hardness ratio map for the energy splitting of $1.0$~\keV{} (S/H = E[0.3-1.0~\keV{}]/E[1.0-10.0~\keV]) {\it Lower:} The hardness ratio map for the energy splitting of $6.0$~\keV{} (S/H = E[0.3-6.0~\keV{}]/E[1.0-10.0~\keV]).}
	\label{fig:HR}
\end{figure}

From the two longest exposures (observation id 6930 and 7289) we have created two hardness ratio maps with different energy splittings between the soft and hard photons. One has a splitting energy of $1.0$~\keV{} so S/H = E[0.3-1.0~\keV{}]/E[1.0-10.0~\keV] and the other of $6.0$~\keV{} so S/H = E[0.3-6.0~\keV{}]/E[6.0-10.0~\keV]. We produced a soft and a hard exposure corrected image with a binning of $5$\arcsec{}/pixel. The images were smoothed with a Gaussian of width $\sigma = 5$\arcsec{}, before the soft images were divided by the hard images. For visualisation, the hardness ratio maps were smoothed again with Gaussian of width $\sigma = 15$\arcsec{}, resulting in the two hardness ratio maps shown in \figref{HR}. Bright colour means excess of low energy photons. In the case of hydrostatic equilibrium this means colder, but since substructure is rarely in hydrostatic equilibrium, the temperature that can be derived from spectral fitting is not the actual physical temperature.

In the $1.0$~\keV{} hardness ratio map (upper part of \figref{HR}) a  substructure in the NE part of the cluster is clearly visible, where in the $6.0$~\keV{} hardness ratio map (lower part of \figref{HR}) only the (almost) circular symmetric part centered at the X-ray peak is visible. 

The splitting of $1$~\keV{} is commonly used \citep[e.g.][]{Fabian:2000}, but we have investigated several splittings between $0.5$~\keV{} and $7.0$~\keV. For splittings between $1.0$~\keV{} and $6.0$~\keV, the hardness ratio map shifts gradually between the two extremes shown in \figref{HR}, so increasing the splitting energy hides the NE structure and enhances the circular cluster structure.

The contours of the total X-ray emission from Abell~1689 (green in \figref{HR}) give the impression of being circular with only a slight elongation in the North-South direction. This elongation is more pronounced in the hardness ratio maps with the $1$~\keV{} splitting. 

In the hardness ratio map with splitting at $6.0$~\keV{} there is a relative soft-photon excess centred at the position of the total X-ray peak, which indicates a cool core in the cluster. 

From the hardness ratio maps it is inferred that Abell~1689 consist of a spherically symmetric part, which to the NE is obscured by come substructure. The cool core indicates that the spherical part of Abell~1689 is relaxed and in hydrostatic equilibrium \citep{Voigt:2006}. Therefore we have split the cluster into two regions and in the following sections, the temperature and mass profiles have been determined from the symmetrically appearing SW part alone. The division line shown in \figref{xray} and \figref{HR} (white dashed) goes through the X-ray peak and is tilted by 20$^o$ with respect to the East-West direction.

\subsection{X-ray analysis and results: Temperature profile} \label{sec:analysis_temp}
Assuming the intra cluster medium to be an optically thin and completely ionised gas, we determined the global properties of Abell~1689 by analysing a $0.5-8.0$~\keV{} spectrum of the central $3$\arcmin. This radius was chosen because it includes most of the cluster emission (the dashed green circle in \figref{xray}) and the same radius is used in earlier analyses \citep{Andersson:2004,Lemze:2007}. An isothermal plasma model (MEKAL, \citet{mekal}) including Galactic absorption was fitted to the spectrum using the spectral fitting package Xspec 12.3 \citep{xspec}.  The absorption in neutral hydrogen along the line of sight was fixed to $1.83\times10^{20}\cm^{-2}$ \citep[$n_H$ tool]{n_H}. Leaving $n_H$ a free parameter while fitting gave a consistent value. The redshift was fixed to $z=0.183$. The redshift can also be determined from the spectral fitting, but the result had an uncertainty of $\approx 20$\% and as a consequence we used the fixed value. Unfortunately this also excludes the idea of determining the distance along the line of sight to the NE substructure from the spectral fitting.

We obtained a global temperature of $10.5\pm 0.1$~\keV{} and an abundance of $0.37\pm0.02$ solar value for a reduced $\chi^2$ of 1.3 (1725 degrees of freedom). This is consistent with earlier reported results \citep{Xue:2002,Andersson:2004,Lemze:2007}. 

Clusters are not isothermal and consequently we made a radial temperature analysis. The X-ray emission of the NE part of Abell~1689 contains substructure in the low energies (see \secref{result_HR} and \secref{discussion_xray}) indicating that this part of the cluster is complex. Hence, we have restricted the radial temperature analysis to the SW symmetrically appearing half of the cluster.

\begin{figure}[tbp]
\centering
	\includegraphics[width=8.5cm]{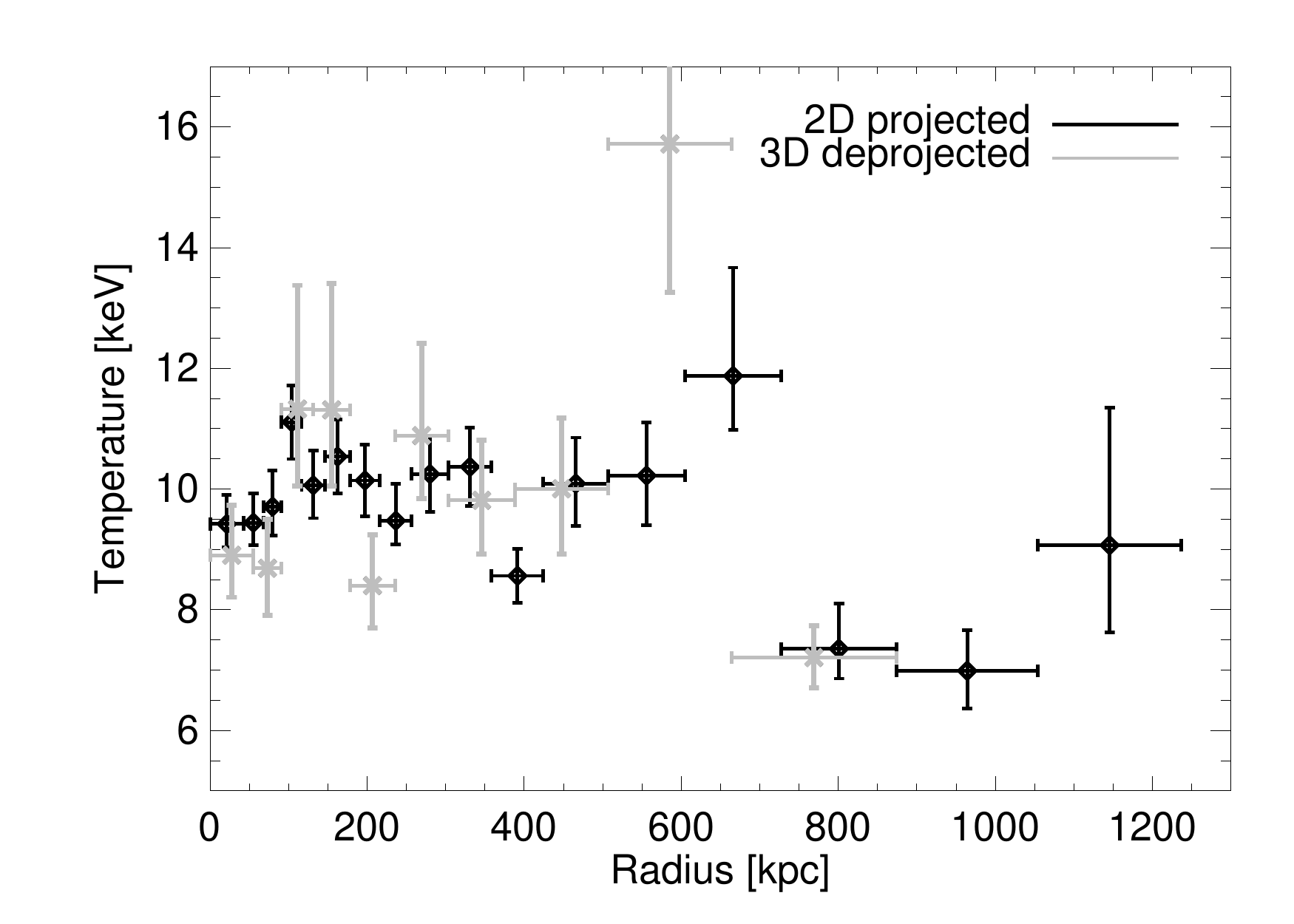}
	\caption{The temperature profile reconstructed from the SW symmetric part of Abell~1689. The black diamonds are the 2D projected profile and the grey crosses are the 3D deprojected profile. For the further analysis the 2D projected profiles was used.}
	\label{fig:temp}
\end{figure}

Using the four data sets and the blank sky background, we have extracted the spectra from the SW half of Abell~1689 in 17 half-circular bins centered at the X-ray peak and containing at least 30000 events each (green circles in \figref{xray}). The same model as for the global properties was fitted to the $0.5-8.0$~\keV{} spectrum of each radial bin. The obtained (projected, 2D) temperature profile is shown in \figref{temp} (black diamonds). 

Fitting the complete annuli provides similar temperatures except for the annuli affected by the substructures. In general the reduced $\chi^2$ of the fits are closer to one for the half annuli than for the complete annuli.

Since the observed X-ray emission is a superposition of gas shells of different temperatures we proceeded to deproject the temperature profile of Abell~1689 using ten half-circular bins, each containing at least 45000 counts. Using Xspec, the spectra and response matrices for all bins were fitted simultaneously to the same plasma model as for the 2D temperature profile, but now combined with the {\it projct} command, which accounts for the projection effects performing a 3D to 2D projection of prolate shell annuli. Each shell are assumed to have its own temperature and all the temperatures are allowed to vary independently. Spherical symmetry of the SW part was assumed as supported by the hardness ratio maps. The deprojected (3D) temperature profile is shown in \figref{temp} as grey crosses. 

The two temperature profiles are quite similar and for that reason, we have reconstructed the mass profile from the 2D projected temperature profile.

\subsection{X-ray analysis and results: Mass profile}
\begin{figure}[tbp]
\centering
	\includegraphics[width=8.5cm]{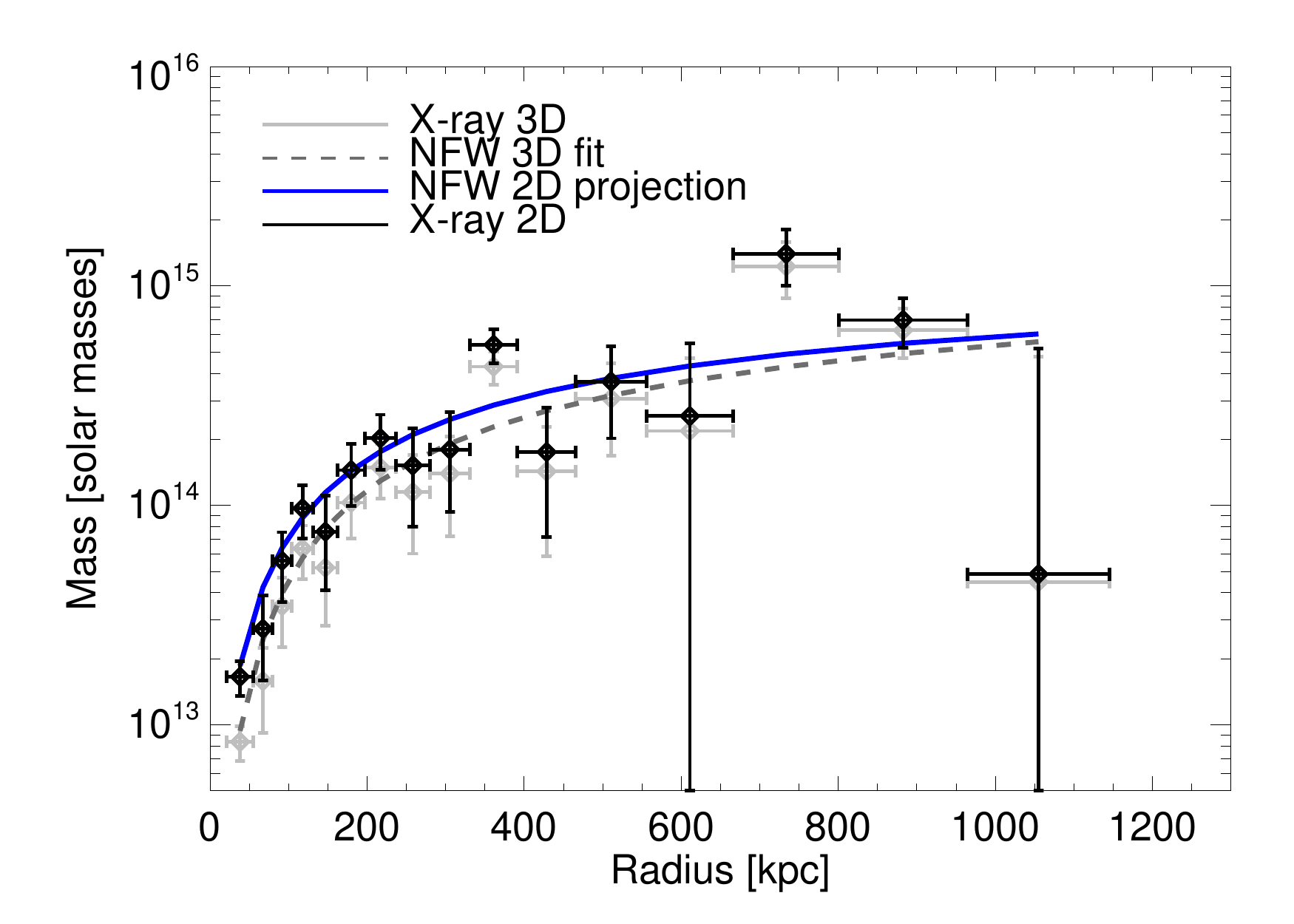}
	\caption{The 3D X-ray mass profile (grey diamonds). The fitted NFW profile is shown as a grey dashed line, and the projected 2D NFW profile as a blue solid line. The 2D projected X-ray mass profile i shown as black diamonds.}
	\label{fig:mass_project}
\end{figure}

Assuming the SW part of Abell~1689 to be one half of a spherical symmetric cluster in hydrostatic equilibrium, the 2D projected temperatures can be used to determine a mass profile following the procedure described in \citet{Voigt:2006}. The method is based on the hydrostatic mass equation \citep{Sarazin}:
\begin{equation}
M_{3D}(<r) = -\frac{k_b T_g(r) r}{G\mu m_p} \left(\frac{d \ln (T_g(r)) }{d \ln (r)} + \frac{d \ln (\rho_g(r)) }{d \ln (r)} \right) \, , 
\end{equation}
where $k_B$ is the Boltzmann constant, $G$ is the gravitational constant, $\mu$ is the mean particle weight weighted by the proton mass $m_p$. $T_g$ is the gas temperature as a function of radius, and $\rho_g$ is the gas density as function of radius, which can be found from the normalisation of the fitted plasma model. The resulting 3D mass profile of the SW part is shown in \figref{mass_project} (black). The errors bars are determined by standard error propagation.

\subsection{X-ray analysis and results: NFW fit and projection} \label{sec:analysis_projection}
The mass profile determined from X-rays describes the 3D matter distribution. However the mass profiles from gravitational lensing (described in \secref{lensing}) are 2D projected profiles, so in order to compare the two profiles, we need to project the X-ray profile as well. It is most easily done by projecting the density along line of sight, and then integrate along the projected radius, $R$ \citep{Sarazin}:
\begin{equation}\label{eqn:projection}
M_{2D}(<R) =  \int_{0}^{R} 2\pi R' \int_{R'}^{\infty} \frac{\rho_{tot}(r) r}{\sqrt{r^2-R'^2}}dr  dR' \, .
\end{equation}

If we naively calculate the density profile from the discrete X-ray mass profile, it becomes unphysical with negative densities because of the fluctuations in the mass (even if the uncertainties are taken into account). Instead we have investigated how large the projection effect is for different analytical density profiles. The projection mainly affects the inner part of the mass profile by a factor of up to two, and leave the outer parts effectively unchanged. In the very central parts of the cluster, the projection depends on the cuspiness of the chosen profile. However, the X-ray mass profile does not probe the central part in great detail and the chosen profile is required to reproduce the observed 3D mass profile by fitting, so we have chosen to use an NFW profile \citep{NFW}. The 3D NFW integrated mass profile was fitted to the 3D X-ray mass profile using reduced $\chi^2$ statistics. The best fit is shown as a dashed grey line in \figref{mass_project} and has the following parameters: $r_s=174\pm 10$~\kpc{}, $\rho _0 = (7.79\pm 0.02) \times 10^6 \Ms$/\kpc$^{3}$, $c_{200}=5.6$ for a reduced $\chi^2=1.6$ (13 degrees of freedom) consistent with earlier published results \citep{Lemze:2007}. The NFW profile was projected using \eqnref{projection} and the obtained 2D profile was divided by the 3D profile to achieve a radius-dependent projection factor, which ranged from approx. two in the centre of the cluster to one in the the outer parts. Each point in the 3D X-ray mass profile was converted to a 2D mass profile by multiplying with the projection factor at the corresponding radius. \figref{mass_project} shows the 3D profile as grey diamonds, the fitted NFW profile as a grey dashed line, the projected 2D NFW profile as a blue solid line, and the corrected 2D X-ray mass profile as black diamonds. The one $\sigma$ error bars were correspondingly corrected.

\subsection{X-ray analysis and results: The NE region} \label{sec:analysis_NE}
The NE region is dominated by soft photons (see \secref{result_HR}) and for that reason expected to be colder than the SW region. Fitting a single temperature component plasma model (same model as above) to the spectrum of the region within the white circle in \figref{HR} gives a temperature of $9.3 \pm 0.9$~\keV{} (reduced $\chi^2 = 0.97$ for $129$ d.o.f.). This is slightly colder than the corresponding SW part of the cluster, which has a temperature of $11.1 \pm 0.6$~\keV{} (reduced $\chi^2 = 1.02$ for $707$ d.o.f.) at similar distance to the centre. The emission from the region is assumed to be a combination of the emission from the SW spherical component plus some softer component.

We have attempted to determine the temperature of the NE component, by fitting a two temperature model to the spectrum of the NE region. One of the temperatures was fixed to the temperature of the spherical component at the same distance ($11.1$~\keV) while the other temperature was a free parameter of the fit. We obtained a cold component temperature of $1.3 \pm 1.0$~\keV{} (reduced $\chi^2 = 0.92$ for $128$ d.o.f.) with a normalisation of 9\% of the normalisation of the warm component. \citet{Mazzotta:2004} demonstrates that in general it is very hard to disentangle two temperature components with equal normalisation unless in the case where one temperature is around $10$~\keV{} and the other one well below a few \keV{}. Comparing to our scenario, the normalisations are not equal. Therefore we have tested the stability of the two temperature fit by varying the two temperatures and their fractional normalisation and then fitting for the overall normalisation. This procedure gives similar reduced $\chi^2$ for a range of parameters. For a reduced $\chi^2<1.5$ the two temperatures can be varied independently between $4$~\keV{} and $14$~\keV{} as long as the normalisations stays a free parameter. The conclusion is that the best fit for the NE region is either a two temperature model with a warm and a very cold component or a one temperature model with a temperature slightly lower than the corresponding SW part of the cluster, which is consistent with fitting a one temperature model to a two temperature plasma \citep{Mazzotta:2004}.

\subsection{Discussion of X-ray results} \label{sec:discussion_xray}
The hardness ratio maps and spectral analysis reveal that Abell~1689 is not spherically symmetric as indicated by the total X-ray image, but rather consists of a spherical main clump centred at the X-ray peak and a softer X-ray emitting substructure to the NE. 

The NE substructure is not easily visible in the total X-ray emission. It is because the total cluster X-ray emission is generally dominated by low energy photons almost regardless of gas temperature, which again is dominated by emission from the main cluster. The absolute number of photons leading to the relative difference in the soft to hard photon ratio is simply drowned in the total X-ray emission. However, the substructure is clearly visible in the ratio between the surface brightness profiles of the two halves shown in \figref{ratio}.

The substructure visible in the hardness ratio in the upper part of Abell~1689 is consistent with the results from \XMM{} presented by \citet{Andersson:2004}, but it is more distinct in the work presented here.

The symmetric SW part of Abell~1689 has a cool core indicated by the hardness ratio map and by the temperature profile. The cool core has earlier been claimed missing in Abell~1689 \citep{Andersson:2004}. The cool core indicates, that the SW part of Abell~1689 is in hydrostatic equilibrium \citep{Voigt:2006}. 

\begin{figure}[tbp]
\centering
	\includegraphics[width=8.5cm]{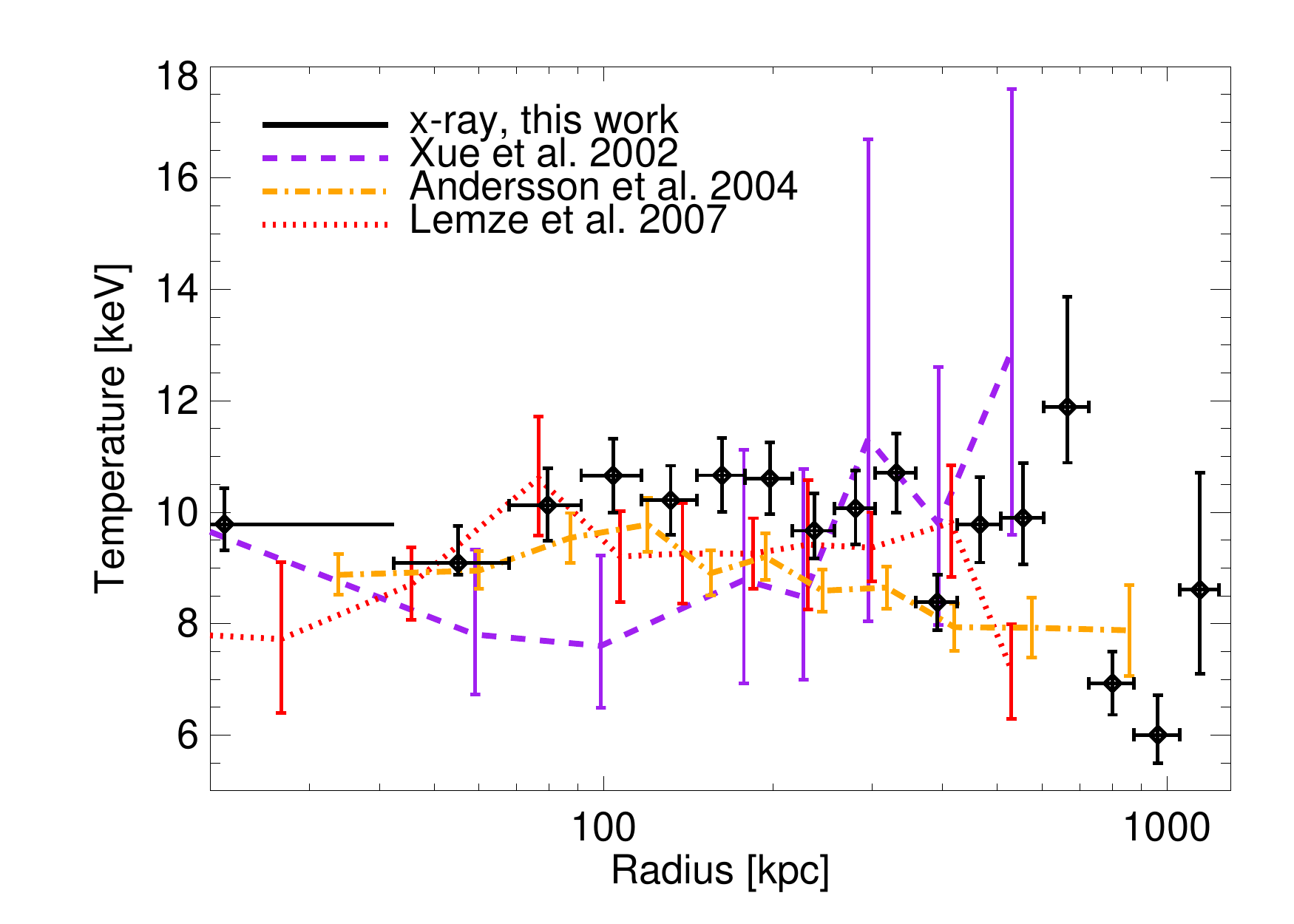}
	\caption{The obtained temperature profile compared to earlier results: purple (dashed) from \citet{Xue:2002}, orange (dashed dotted) from \citet{Andersson:2004} (3D deprojected), and red (dotted) from \citet{Lemze:2007}.}
	\label{fig:xraycomp}
\end{figure}

The temperature profile obtained from the SW part of the cluster is compared to earlier published X-ray temperature profiles in \figref{xraycomp}. The error bars are smaller than earlier obtained profiles from \Chandra{} data \citep{Xue:2002,Lemze:2007} due to better statistics, and the measurements extend to larger radii due to the use of blank sky background. Only the analysis of \XMM{} data by \citet{Andersson:2004} has comparable statistics and extension.  However the temperature profile obtained in this work is generally slightly warmer which leads to an overall larger mass. 

The X-ray mass profile as inferred from the SW part of the cluster is well fitted by an NFW profile with a concentration parameter of $c_{200}=5.6$. This values is in better agreement with the value of $c_{200}=4.7\pm1.2$ predicted by \citet{Neto} from cosmological simulations, than the value of $c_{200}=7.6\pm1.6$ obtained from gravitational lensing \citep{Limousin:2007} or the value of  $c_{200}=7.7$ from earlier X-ray analyses \cite{Andersson:2004}. This supports our interpretation of the SW part being spherical and in hydrostatic equilibrium.

\section{Additional observations of Abell~1689}
\subsection{Strong and weak lensing analyses} \label{sec:lensing}
We compare our X-ray analysis to strong and weak gravitational lensing analyses of Abell~1689. Strong and weak gravitational lensing are two different methods of determining the mass of the lensing object(s) based on distortions of the images of background sources. Strong lensing is tracing the mass of the lens within the Einstein radius where arcs and multiple images are forming. Weak lensing traces the mass outside the Einstein radius where the gravitational force is weaker, and stretches the images of background objects perpendicular to the direction to the center of the lens. Thus the two methods are valid in different ranges, overlapping only around the Einstein radius.

\begin{figure}[tbp]
\centering
	\includegraphics[width=8.5cm]{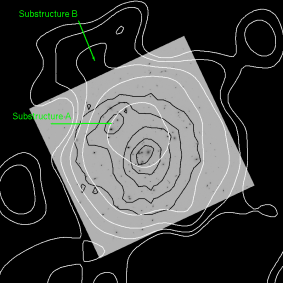}
	\caption{Optical images (HST/ACS) with gravitational lensing contours. The black contours are the mass contours from strong lensing. The minimum value is $10^{9}\Ms / \arcsec ^2$ and the distance between contours is $3.5 \times 10^{8}\Ms/ \arcsec^2$. The strong lensing has a mass peak at the centre of the cluster and a smaller peak $\approx 180$~\kpc{} to the NE (A). The white contours are the $\kappa$ contours from weak gravitational lensing with a resolution of $50\arcsec$ (FWHM). The minimum value is $\kappa = 0.1$ and the distance between contours is $\delta \kappa = 0.16$. The weak lensing contours clearly shows a subclump $\approx 500$~\kpc{} to the NE of centre (B).}
	\label{fig:optical}
\end{figure}

There have been many strong lensing studies of Abell~1689 \citep{Broadhurst:2004,Broadhurst:2005,Limousin:2007,Lemze:2007}. Here we have used the latest and most detailed by \citet{Limousin:2007}, which is based on data from the HST Advanced Camera for Surveys (ACS) with the spectroscopic information of the lensed objects in the system from the Keck Telescope and the Very Large Telescope (VLT). The mass reconstruction was done using a Bayesian Monte Carlo Markov Chain method with LENSTOOL \citep{Jullo:2007}. The result is an accurate mass mapping of the central parts of Abell~1689, which demonstrates a clear bimodality as shown in \figref{optical} (black contours). In the very centre there is a main clump associated with the peak of the X-ray emission and then a second significant clump $\approx 180$~\kpc{} NE of the centre (A in \figref{optical}) associated with a clump of galaxies (not to be confused with the X-ray structure to the NE, at a distance from $\approx 500$~\kpc{} to the centre). Previous studies \citep{Broadhurst:2005,Halkola:2006b} were using fewer spectroscopically confirmed imaged background systems but are in good agreement with the results from \citet{Limousin:2007}. 

\begin{figure}[tbp]
\centering
	\includegraphics[width=8.5cm]{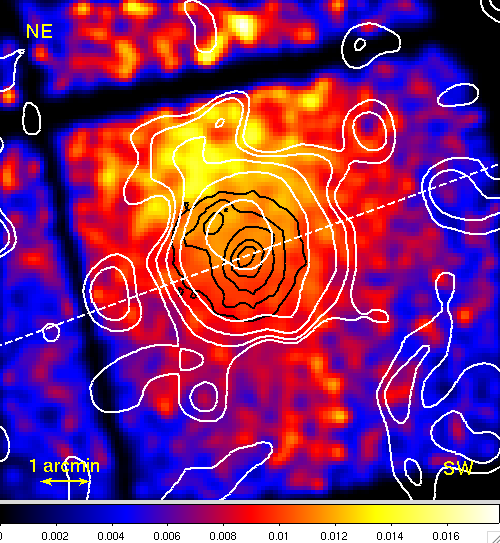}
	\caption{The hardness ratio maps with energy splitting of $1.0$~\keV{} (S/H = E[0.3-1.0~\keV{}]/E[1.0-10.0~\keV]). Bright is soft photon excess. The white line divides the cluster into the NE (upper) part and the SW (lower) symmetric part. The contours are identical to \figref{optical} with white being weak gravitational lensing and black strong gravitational lensing.}
	\label{fig:HR_lensing}
\end{figure}

For weak gravitational lensing we compare to a recent analysis by \citet{Dahle:2008}, which was based on a mosaic of 16 HST WFPC2 pointings covering the central $\sim1.8~\Mpc{} \times 1.4~\Mpc{}$ of the cluster, complemented at larger radii by ground-based data using the CFH12K mosaic CCD camera at the 3.6~m Canada-France-Hawaii Telescope (CFHT), covering a larger field of $7.7~\Mpc{} \times 5.2~\Mpc{}$. This provides a unique mass map with an unprecedented combination of spatial resolution ($50\arcsec$) and a large spatial extension. The surface mass density map from weak gravitational lensing shown in \figref{optical} (white) was reconstructed by \citet{Dahle:2008} using the \citet{Kaiser:1993} method. This method relies on the fact that the convergence and the two components of the shear are linear combinations of the second derivative of the effective lensing potential. Using Fourier transformation of the convergence and the shear, one can obtain linear relations between the transformed components and determine the mass. The mass contours clearly shows a non-spherical morphology of Abell~1689 at $\approx 500$~\kpc{} from the centre (B in \figref{HR_lensing}).

\subsection{Lensing analysis: mass profiles} \label{sec:analysis_lensing}
From weak lensing the one dimensional radial mass profile was obtained using aperture mass densitometry \citep{Kaiser:1995,Clowe:2000,Fahlman:1994}. This method determines the mean surface mass density within an aperture minus the mean density in a surrounding annulus. Hence, a lower limit to the projected mass of the lens inside the aperture can be found. The degree of underestimation of the true enclosed mass depends on the inner and outer radii and the slope of the projected mass density profile. For realistic projected density profiles (steeper than $r^{-1}$ at large radii) the mass results presented here will be underestimated by less than 30\% at all radii.

For strong gravitational lensing the mass map was integrated in radial bins of 1\arcsec{} centered on the peak of the X-ray emission (which coincides with the strong lensing mass peak).

For both weak and strong lensing we reconstructed two different radial mass profiles: one by integrating the whole surface density map and the other by integrating the SW half of the surface density map as in the X-ray analysis (but multiplied by two to get a complete sphere).

\subsection{Lensing results and discussion} \label{sec:result_lensing}

Both strong and weak lensing analyses show that Abell~1689 is not spherical, but has substructure in the NE direction on both small ($\approx 180$~\kpc{}) and large scales ($\approx 500$~\kpc{}).

\begin{figure}[tbp]
\centering
	\includegraphics[width=8.5cm]{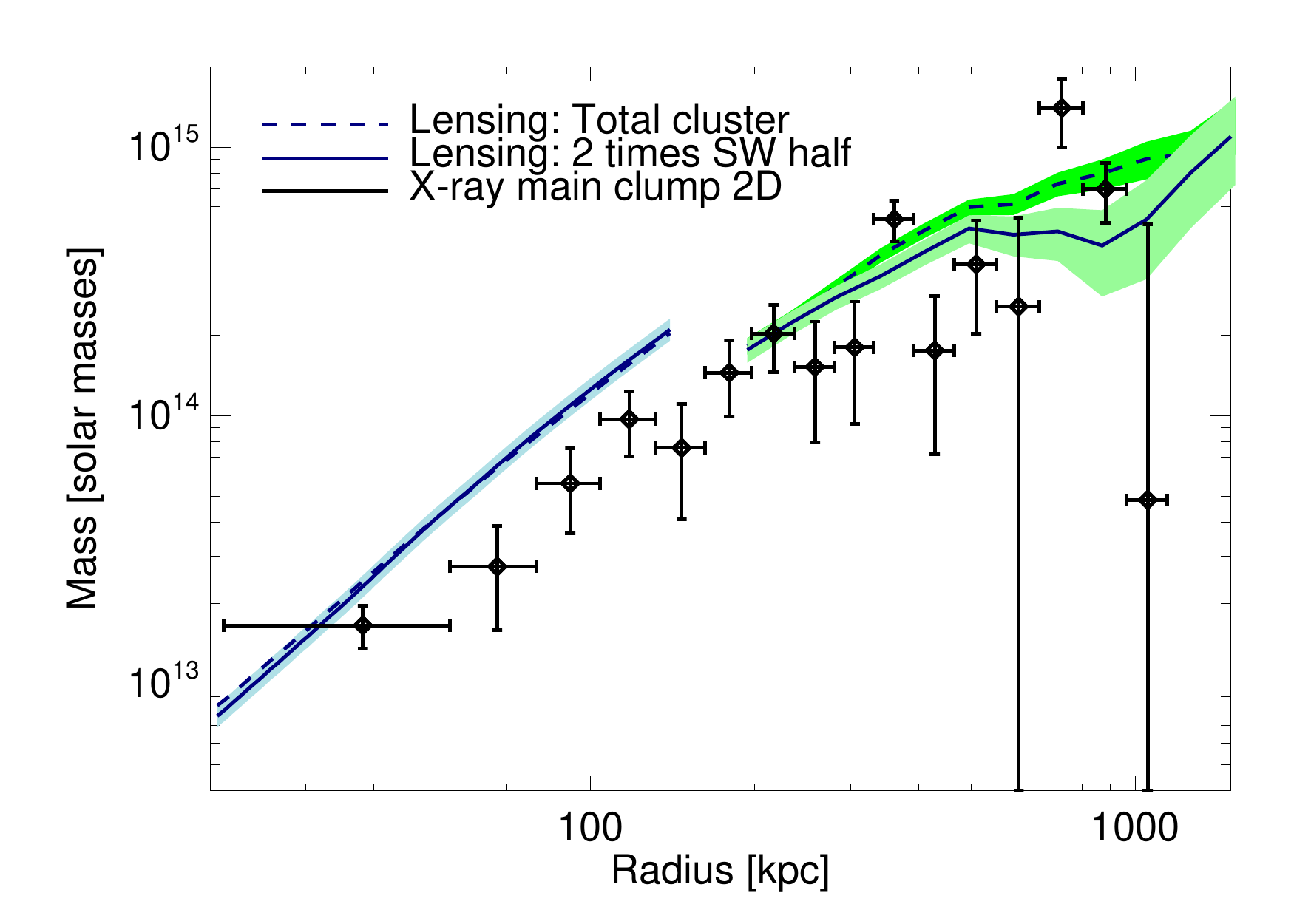}
	\caption{The 2D projected mass profile reconstructed from X-ray (black), strong gravitational lensing (blue) and weak gravitational lensing (green). The dashed lensing profile is for integration over the full cluster and the lower solid curve is for integration over the SW part multiplied by two (in order to get a full sphere). The width of the shaded regions corresponds to the one sigma errors on the lensing profiles.}
	\label{fig:mass}
\end{figure}

The mass profiles of the SW part of Abell~1689 obtained from gravitational lensing are shown in \figref{mass}. The mass profiles were plotted keeping in mind that the strong lensing mass map is valid within the Einstein radius, which for Abell~1689 is $45$~\arcsec$\approx139$~\kpc{}. The dashed curves in \figref{mass} are the integration of the whole mass map and the solid curves are the integration of the SW part only, but multiplied by a factor of two to get a full sphere. 

For the strong lensing, the mass profile integrated over the total cluster is very similar to the mass obtained by integrating the SW part only since the NE substructure at $\approx 180$~\kpc{} is negligible compared to the total mass of the cluster. For the weak lensing, the difference is a bit smaller and the total integration gives a larger value due to the NE substructure also seen in the X-rays. The total mass within 1200~\kpc{} is $\approx 9\times10^{14}~\Ms$. 

The mass of the weak lensing substructure is determined to be $1.25 \pm 0.3 \times 10^{14}\Ms$ within $1.56~\arcmin$.
 
\begin{figure}[tbp]
\centering
	\includegraphics[width=8.5cm]{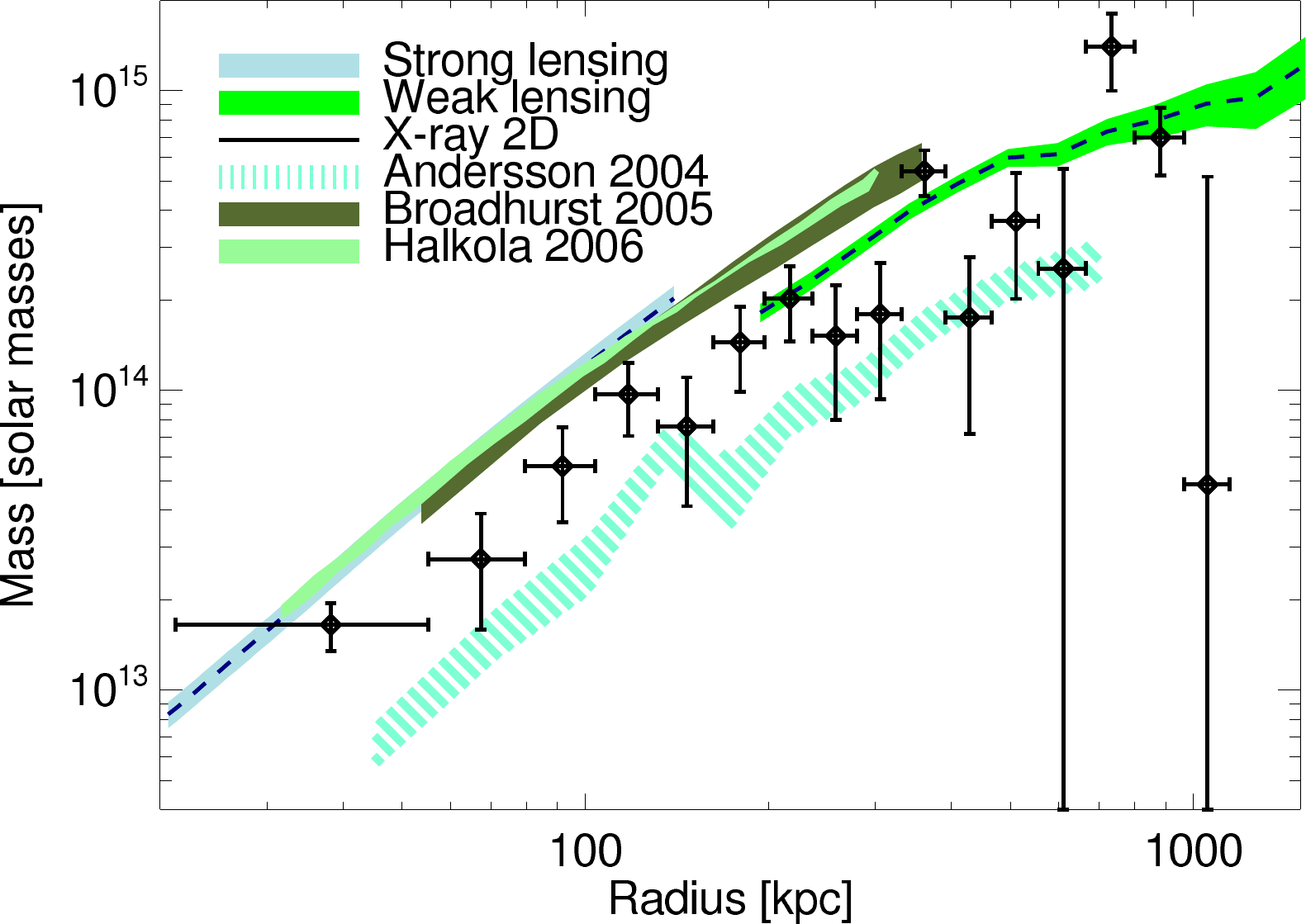}
	\caption{The weak and strong gravitational mass profiles (dashed navy) of the total cluster compared to earlier results by \citet{Broadhurst:2004} (olive green) and \citet{Halkola:2007} (green).  Also the X-ray mass profile obtained here is shown (black crosses) together with the one obtained by \citet{Andersson:2004} (aquamarine dashed).}
	\label{fig:masscomp}
\end{figure}

In \figref{masscomp} the total cluster mass profile is compared to earlier studies of Abell~1689 with weak gravitational lensing. It is seen that there is good agreement between the strong lensing mass profiles for radii smaller than 100~\kpc. Above 100~\kpc, the profile of \citet{Dahle:2008} is slightly lower than the earlier profiles \citep{Broadhurst:2004,Halkola:2007,Umetsu:2007}. 

Assuming the strong lensing mass profile to be the actual 2D projected mass profile of the inner part of Abell~1689 and the X-ray profile to be the actual 3D profile allows us to constrain the 3D to 2D projection factor as a function of radius, which depends on the cuspiness of the mass profile. The maximally allowed projection factors are $M_{2D}/M_{3D} = (2.8, 4.2, 3.1, 2.6)$ for radii of $r=(38, 67, 92, 188)$~\kpc{}, which are all larger than the factors determined from the NFW profile projection in \secref{analysis_projection}.

\subsection{Redshift distribution of cluster galaxies}
The redshift distribution of the galaxies within Abell~1689 is overall Gaussian \citep{Duc,Teague:1990,Czoskephd}, but with irregularities \citep{Lokas:2005}. The full width at half maximum of the main structure is $\Delta z = 0.02$ corresponding to a line of sight velocity dispersion of $\approx 2500$~km/s. This is in contrast to other clusters where very few rich clusters have been found with a velocity dispersion above $1000-1200$~km/s. It suggests either that Abell~1689 is not in hydrostatic equilibrium or it is a structure with several clumps along the line of sight. The latter interpretation is favoured by \citet{Girardi:1997} and by \citet{Lokas:2005} who compared the velocity distribution of Abell~1689 to simulations and argued that the high velocity dispersion is due to several clumps along the line of sight. Removing what they found to be separate unbound structures, they got a total dynamical mass of $2\times10^{15}\Ms$ within $3.5$~\Mpc.

Also \cite{Lemze:2008} have studied the dynamics of Abell~1689. They derived the virial mass of the cluster to be $1.73 \pm 0.59 \times 10^{15}\Ms$ using the velocity caustics of the cluster galaxies.

\section{Discussion}
Comparing the X-ray hardness ratio maps with the weak and strong gravitational lensing mass maps (\figref{HR_lensing}) we see a coincidence between the substructure in the X-ray emission and the bimodality in the lensing maps. The total X-ray emission (\figref{xray}) features a slight elongation along the same axis as the secondary clump in the strong lensing map. Further from the centre, the position of the NE substructure in the hardness ratio map coincides with the position of a large substructure in the weak lensing map. On the opposite site, the SW part of the cluster appears circular and symmetric in all observations. From this we conclude that Abell~1689 consist of a spherical main clump and some substructure to the NE, which is interfering with the emission from the NE half of the main clump.

The NE half of the cluster is slightly brighter than the SW half at the distance of the weak lensing substructure (\figref{ratio}).

The mass of the substructure is $1.25\pm0.3\times10^{14}\Ms$ within a radius of $1.56$~\arcmin determined from weak lensing. This is only a small fraction of the total cluster mass. The gas mass within the same radius has been estimated assuming sphericity and uniform gas density. The result of $\approx 9.7\pm0.3\times10^{12} \Ms$ only contributes with a very small fraction to the total cluster mass. Nonetheless the X-ray emission from the substructure is bright enough to introduce a complication of the spectral analysis and thereby potentially introduce a mis-interpretation of the temperature of the cluster.

Reconstructing mass profiles from the SW half of the main clump only leads to unprecedentedly good agreement between weak and strong gravitational lensing, and the 2D projected X-ray mass profile, which supports the interpretation of Abell~1689 consisting of a spherical main clump and substructure to the NE.

We have quantified the deviation between the different mass profiles by comparing the relevant part of the 2D projected X-ray mass profile to the weak and strong lensing profiles separately. The values of one profile were compared to a linear interpolation of the other profile. The reduced $\chi^2$ was determined to be 1.8 (four degrees of freedom) for the weak lensing and X-ray comparison and 1.2 (10 degrees of freedom) for comparison of the strong lensing to the X-ray mass profile.

The cool core indicates that Abell~1689 is not a recent merger and supports the claim that the SW main clump is in hydrostatic equilibrium. It indicates that the NE substructure is not leftovers from an early collision and maybe not gravitationally interacting with the main clump at all. If it is infalling, the merging is in such an early state that the hydrostatic equilibrium of the SW part has not yet been disturbed. The line of sight galaxy velocity distribution supports a scenario with several clumps along the line of sight, where the gas does not interact.

The hardness ratio maps can be used as easy diagnostics for the thermal distribution of the cluster, which is related to any substructure. For Abell~1689 the substructure to the NE is visible in a hardness ratio map with an energy splitting of $1.0$~\keV{} and a resolution of $10$\arcsec{} made from $15~\ks$ of the 6930 observation.  Many cluster observations have exposures of the order of $15~\ks$ so we propose to use hardness ratio maps in the selection process of relaxed clusters to be further studied.

In this work we have taken advantage of high quality X-ray and gravitational lensing data, which have improved the agreement between several mass estimation methods significantly.

\section{Conclusions}
The new X-ray data analysed here and compared with weak and strong gravitational lensing data show that on large scale Abell~1689 appears spherical, in agreement with earlier data. However, both temperature structure and gravitational lensing shows that Abell~1689 contains some substructure to the NE. The main part appears circular and is centered at the peak of the total X-ray emission. Only the SW half is not significantly influenced by other structures. From X-ray hardness ratio maps and temperature profile it is seen that the SW part features a cool core. 
The NE substructure is seen in the weak gravitational lensing mass map and the X-ray hardness ratio maps. It has an excess emission of soft X-ray photons relative to hard photons.

We have determined the mass profile of the spherical main clump of the cluster from X-ray observations and compared to recent gravitational lensing results. The obtained cluster mass within $\approx875$~\kpc{} derived from X-rays alone is $6.4\pm2.1\times10^{14}\Ms$ compared to a weak lensing mass of $8.6\pm3.0\times10^{14}\Ms$ within the same radius. The profiles are in very good agreement out to $1200$~\kpc{} and the discrepancy between X-ray and lensing mass profiles has been significantly reduced due to high quality data.

\acknowledgements
We would like to thank M.~Markevitch for useful comments on the background analysis.
The Dark Cosmology Centre is funded by the Danish National Research Foundation.
KP acknowledges support from Instrument Center for Danish Astrophysics. 
ML acknowledges support from the Agence Nationale de la Recherche (France) project n° BLAN06-3\_135448.
HD acknowledges support from the Research Council of Norway
\bibliographystyle{apj}
\bibliography{references}

\end{document}